# Statistics against irritations: a response to Dickens's apologists

By Mikhail Simkin

In a recent article [1] I reported the results of the test[1], where the takers had to tell the prose of Charles Dickens from the prose of Edward Bulwer-Lytton. The former is a required reading in school, and the latter has a bad writing contest[2] named after him. Nevertheless, the test-takers performed on the level of random guessing. This research got some media attention.

While Mark Howarth's article in *The Daily Mail* [2] is rational, the article by Alison Flood in *The Guardian* [3] is emotional. She even bills her talking points as "irritations:"

"My first irritation is with the assumption that Bulwer-Lytton is the worst writer in history. This is just ludicrous. Yes, there might be the annual Bulwer-Lytton Fiction Contest…"

There might be? I learned about Bulwer-Lytton through this contest. Is it just a peculiarity of my experience? Let us have a look at statistics. I searched my local library catalogue[3] for Bulwer-Lytton and got 31 results, the first three of which were the collections of "the best (?) from the Bulwer-Lytton Contest." A similar search for Dickens produced 1,275 results the first three of which were biographies. Even if we forget the contest and compare the bare numbers, we see that there are forty times more Dickens titles than Bulwer titles in the library. This means that they are writers of distinctly different status. Given just that, are not the test results remarkable?

Another critic, Gina Dalfonzo from *The Atlantic*, wrote [4]:

"First, hardly anyone argues that Edward Bulwer-Lytton was the worst writer of all time. That someone could even think of making that contention in the age of *Twilight* and *50 Shades of Grey* boggles the mind. Mediocre he may have been, but a joke contest inspired by seven words that he wrote cannot stand as the sole proof that he was the most awful author ever."

If nobody argues that Bulwer is the worst writer in history of letters, why has he the worst writer contest (inspired by 58 words that he wrote) named after him? Of course, the badness is in the eye of a beholder and there are other contenders for the title of the worst ever. However, even while denying that Bulwer was the worst in the world, Dalfonzo herself calls him "mediocre." Isn't it revealing that people can't tell Dickens from a mediocre writer?

Dalfonzo continues [4]:

"The method of collecting data seems just as shaky ….So we know we have an educated subset because of their location, even though we have no idea whether the person sitting at any given computer was a professor, a student, or a janitor."

During all those years on campus, I did not see a janitor at a computer. To make the objection plausible the spoilers should have been secretaries. They had caused trouble before. Richard Nixon's secretary erased some tapes and the President had to resign his office. They must be the

---

[1] http://reverent.org/bulwer-dickens.html
[2] http://www.bulwer-lytton.com/
[3] http://inland.librarycatalog.info/polaris/default.aspx

ones to blame for the scandalous results of the quiz as well. But let me give a word to a more diligent commenter, Taylor Malmsheimer from *New York Daily News* [5]:

"Skeptical? We were too, so we asked an English major at Dartmouth — which is allegedly in the Ivy League — to take … [the] quiz…She correctly answered six out of the 12 questions…"

Now let us turn to the second irritation of Alison Flood [3]:

"My second irritation is the assumption that the quality of an author can be judged on an extracted sentence."

I used the passages consisting of several sentences: four, on average. Flood either can't count or used a hyperbole. If the latter is true, it is strange that she did not understand my hyperbole "the worst writer in history of letters," since she must be familiar with the device.

Dalfonzo has similar concerns [4]:

"It's worth noting that only descriptive passages are used; there's virtually nothing involving plot or characterization, even though, as Simkin admits in his paper, these tend to be essential to novels. This omission puts Dickens, known for strong and unique characterization, at a distinct disadvantage."

To learn the plot one has to read the whole novel. To describe a character one needs several pages. Obviously, I could not make the quiz that long, since nobody would take it. I made a short quiz, which compares prose styles. And it is the prose style what they ridicule Bulwer for.

In her *Times Literary Supplement* article about "Bulwer-Lytton, the great unreadable" Joan Sutherland writes [6]:

"Why does he not even have a single title in the 700-strong catalogues of Penguin and Oxford World's Classics? The absurdity of his prose style to modern ears would seem to be a main reason."

I did check the catalogues and found twenty Dickens's titles in Penguin Classics and fifteen in Oxford. The test results show that Dickens's and Bulwer's prose styles are of the same quality. Therefore, Sutherland's answer to her own question is unsatisfactory. Before giving my answer, I'll quote Dalfonzo for the last time:

"Simkin finishes, snarkily: "I began this paper with the question: Are the famous writers different from their obscure colleagues? The answer is: Yes, they have more readers." If an academic can publish that with a straight face, and not grasp that he's just fatally undercut his own argument—why would Dickens have so many more readers if there's no difference between the two?—then perhaps the Victorians weren't the ones who didn't know how to write."

Interestingly, if you look few paragraphs above you will read how Dalfonzo nominates *50 Shades of Grey* for the title of the worst ever in place of Bulwer's literary work. However, the novel is on the top of The New York Times bestsellers list[4]. Remarkably, the large number of readers does not vouch for high prose quality in this case. Why should it do so in the case of Dickens?

---

[4] http://www.bbc.co.uk/news/entertainment-arts-17332129

How some elements of culture can become much more popular than the others even when identical in merit I had explained in the third paragraph of the original article [1]. The culprit is that people copy each other's choices. The articles cited in Ref. [1] develop appropriate mathematical models. The example that I used, the scientific citations, may be too difficult for literary critics. Here I'll use a simpler one. In 2011, 20,153 newborn American babies were named Jacob, while only 237 were named Samson[5]. Thus, the name Jacob is 85 times more popular than the name Samson. Is it intrinsically better? And if it is not one should not be surprised that Dickens has 20 titles in Penguin Classics and Bulwer has none at the time when blind reading test finds no difference between the two.

---

[5] http://www.ssa.gov/OACT/babynames/